\documentclass[12pt]{article}
\usepackage{epsfig}
\usepackage{amsfonts}
\usepackage{amsopn}
\usepackage{amsmath}

\title{
\vskip -50 pt
\begin{flushright}
\normalsize\rm AEI-2012-21 \\
\normalsize\rm NORDITA-2011-25 
\end{flushright}
\vskip 20 pt
Dirac equation for embedded 4-geometries}

\author{
Maciej Trzetrzelewski \thanks{e-mail: maciej.trzetrzelewski@gmail.com} \\ \\
Max-Planck-Institut f\"ur Gravitationsphysik, \\
Albert-Einstein-Institut, \\
M\"uhlenberg 1, D-14476 Potsdam, \\
Germany \\ \\ 
NORDITA,  \\
Roslagstullsbacken 23, 106 91 Stockholm, \\
Sweden}

\begin{document}
\date{}
\maketitle

\abstract{We apply Dirac's square root idea to constraints for embedded 4-geometries swept by a 3-dimensional membrane. The resulting Dirac-like equation is then analyzed for general coordinates as well as for the case of a Friedmann-Robertson-Walker metric for spatially closed geometries.  The problem of the singularity formation at quantum level is addressed.}

\section{Motivation}
One of the most important problems in modern cosmology is the fact that according to the Friedmann model the Universe started with the singularity. It is expected that once the consistent description of General Relativity with Quantum Mechanics is obtained the singularity will be removed. 

In the 60' DeWitt \cite{DeWitt} addressed this problem by assuming that the wave functional satisfying the Wheeler-DeWitt equation should be zero whenever the singularity occurs at classical level. In the case of a Friedmann Universe this implies that the wave function should vanish for the scale factor $a=0$. DeWitt's assumption is an example of one of several boundary conditions (e.g. the no-boundary condition \cite{hh}, the tunneling condition \cite{vilenkin} the symmetric condition \cite{zeh} - see \cite{rev1,rev2} for a review on this subject) which one could set to argue that the singularity can be avoided. More recently, it has been pointed out that this aim can be also achieved by using techniques of loop quantum gravity \cite{bojowald}, by considering mini-superspace models in $D=11$ supergravity \cite{KKN} or in (string theory inspired) brane models where  one addresses this problem from a different perspective in which the Big Bang is identified with brane collisions (for a review see \cite{Jean1,Jean2}). All of  the above approaches, although different in details, represent certain attempts to apply laws of Quantum Mechanics in gravitational systems. It should be noted however that the true solutions of the problem may lie in modifying Quantum Mechanics itself \cite{thooft}. 

A similar singularity formation also takes place in much simpler systems of extended objects forming minimal manifolds embedded in Minkowski spacetime. These objects will in most cases collapse to a point after a finite time (there are exceptions e.g. when they rotate which can prevent the collapse: a rotating closed string on a plane \cite{string,jens1} or a rotating flat torus in $\mathbb{R}^4$ \cite{sasha}). In this paper we firstly discuss minimal manifolds (Section 2) and then proceed to more general geometries (Section 3).

The idea that one can consider spacetime as a 4-manifold $\mathcal{M}$ embedded in some ambient Minkowski space $E$, and therefore to quantize not the metric but the coordinates with give rise to the metric, dates back to late 70' \cite{regge,deser} - see \cite{pavsic1} for a comprehensive list of references about the subject. Solutions of Einstein's equations give automatically the solutions
for equations of motion for embedded coordinates - however the converse is not true \cite{deser}. More recently it has been argued that the equivalence can be obtained by modifying the embedding approach \cite{fad1,fad2,fad3} in a certain way.

Although it is not clear what is the physical meaning of the embedding space (i.e. it is not known if there are extra dimensions) let us note that there are several advantages of this approach. The most crucial used in this paper are related to a clear understanding of
\begin{itemize}
\item the time variable (time on $\mathcal{M}$ is inherited from $E$) and therefore the Hamiltonian formalism
\item the notion of the probability density - events, such as a 3-sphere contraction, take place in $E$. 
\end{itemize}

Here, we would like to develop this idea in a novel context: the method used here is an appropriate generalization of our recent work \cite{mt} (see also earlier attempts  in this direction \cite{ho,stedile})  where we have undertaken  Dirac's idea \cite{Dirac1,Dirac2} to formulate a theory of extended objects with spin. This approach (different from considering supersymmetric version of the bosonic theory) is based on the Dirac's square root procedure applied to the constraints for extended objects. By writing the Dirac equation for embedded spacetime, in this way, one necessarily assumes the existence of the spinor field (actually a functional)  of the manifold considered. Therefore the wave function of the embedded Universe will not be a scalar but a spinor. Intuitively, this change should make a major difference - after all bosons "like" to be in the same state while fermions quite the contrary. Therefore if one considers a contracting fermionic 3-sphere, one should observe a repulsion effect when the radius of a sphere is sufficiently small \footnote{We are not claiming, by this argument, that  fermionic condensates should not exist. Of course they do exist in Nature e.g pairs of fermions (Cooper pairs) in condense matter systems \cite{bcs1,bcs2,bcs3} which give rise to a bosonic states and therefore can condensate. However the physical context in this paper is different. We consider a spinorial functional defined on the 3-manifold and study its behaviour when the manifold considered shrinks to a point.}. 

A behaviour of this type we find in the case of FRW metric for the action containing the Einstein-Hilbert term and the cosmological constant term. Equations of motion for embedding coordinates are more general then Einstein's equation for the metric -  in particular they allow for singular solutions (at $a=0$) in situations where Einstein's equations develop no singularities. We find that at quantum level, when considering the Dirac equation, these singularities are avoided by showing the the wave function is zero for $a=0$.

\section{Minimal 4-manifolds}
Embedded 4-manifold  $\mathcal{M}$ will be described here by coordinates 
$X^A(x)$, $A=0,\ldots,D-1$ where $x^{\mu}$, $\mu=0,1,2,3$, is the internal 
parametrization of the manifold. The metric of the manifold is induced from 
$E$ by 
\begin{equation} \label{embed}
g_{\mu\nu}(x) =\eta_{AB} \partial_{\mu}X^A(x)\partial_{\nu}X^B(x)
\end{equation}
 where $\eta_{AB}$ is a Minkowski metric (throughout the whole paper we are using  $c=\hbar=1$ units as well as the conventions of the Landau-Lifshitz textbook \cite{Landau}, in particular the signature of $g_{\mu\nu}$ is $(+1,-1,-1,-1)$ and correspondingly $\eta_{00}=1$). Such embedding can be found globally for many important geometries which 
are the solutions of EinsteinÕs equations \cite{rosen}. Moreover, it is well known that locally one can always find such embedding in $10$ dimensional Minkowski space \cite{theorem}.

 For the action describing the dynamics of $X^A$ we take the Dirac action \cite{Dirac1,Dirac2,Dirac3} which, for three dimensional membranes, takes the form of the cosmological constant 
term 
\[
S=\int_{\mathcal{M}} \mathcal{L}_{\lambda}d^4x,\ \ \ \ \mathcal{L}_{\lambda} = -\lambda  \sqrt{-g}, \ \ \ g:=\det g_{\mu\nu}.
\]
Here $\lambda$ is a tension and is positive - we will discuss negative tension later on. Varying $S$ with respect to $X^A$ 
gives 
\begin{equation} \label{eom1}
\partial_{\mu}\left(\sqrt{-g}g^{\mu\nu}\partial_{\nu}X^A\right)=0.
\end{equation}
In addition to the equations of motion, there are constraints
\begin{equation} \label{cons1}
\mathcal{P}_A\mathcal{P}^A=-\lambda^2 \det g_{rs}, \ \ r, s = 1, 2, 3
\end{equation}
and $\mathcal{P}_A \partial_r X^A=0$ satisfied by the canonical momenta 
\begin{equation}  \label{momenta1}
\mathcal{P}_A := \partial \mathcal{L}_{\lambda} /\partial(\partial_0X^A) =-\lambda\sqrt{-g}g^{0\mu}\partial_{\mu}X_A. 
\end{equation}

\subsection{Dirac equation}
Equation (\ref{cons1}) is the counterpart of the mass shell constraint for point-like particles. Therefore its linear (in $\mathcal{P}_A$) form will serve as a Dirac-like equation for extended objects. The square root of the l.h.s. of (\ref{cons1}) can be easily performed with use of the Dirac gamma matrices $\gamma^A$ in $D$ dimensions. As for the r.h.s.  there exist at least two ways of doing that. Using the identity
\[
\det g_{rs} = \frac{1}{3!} \{X^A,X^B,X^C\}\{X_A,X_B,X_C\}
\]
and a related one
\[
M^2 = \det g_{rs}\bold{1}, \ \ \ M=  \frac{1}{3!} \gamma^{ABC}\{X_A,X_B,X_C\}
\]
where $\gamma^{ABC}=\frac{1}{3}(\gamma^{AB}\gamma^C+cycl.)$, $\gamma^{AB}=\frac{1}{2}[\gamma^A,\gamma^B]$ and where we used the Nambu bracket $\{f,g,h\}=\epsilon^{rst}\partial_r f\partial_s g\partial_t h$, \cite{nambu}, we find that the constraint (\ref{cons1}) can be linearized as
\begin{equation} \label{sq1}
\left(  -i \gamma^A \frac{\delta }{\delta X^A} + \sqrt{ -\frac{1}{3!}\lambda^2 \{X^A,X^B,X^C\}\{X_A,X_B,X_C\} }  \right) \Psi =0
\end{equation}
or
\begin{equation}  \label{sq2}
\left(  -i \gamma^A \frac{\delta }{\delta X^A} + \frac{1}{3!}i \lambda \gamma^{ABC}\{X_A,X_B,X_C\}   \right) \Psi =0
\end{equation}
where we substituted the functional derivative $\mathcal{P}_A = -i \delta/\delta X^A$. Let us note that there is no ordering ambiguity related to functional derivatives which would have taken place if the equation was second order in $\mathcal{P}_A$ (this is the case in e.g. the Wheeler-DeWitt equation \cite{rev1}). To proceed further a quantum counterpart of the Nambu bracket is required. In the case of membranes where instead of a Nambu bracket there is a Poisson bracket, one uses the matrix regularization \cite{hoppephd} and the Poisson bracket is then replaced by $-i$ times the commutator of $SU(N)$ matrices. Here one faces the problem of regularizing the volumes \cite{volreg}. 

Equation (\ref{sq2}) seems more elegant due to the lack of the square root. Moreover, let us observe that it can be solved exactly by
\[
\Psi = e^S\Psi_0 , \ \ \ S=\frac{\lambda}{24b}\gamma^{ABCD}\int X_A\{X_B,X_C,X_D\}d^4x
\]
where $\Psi_0$ is a constant spinor, $b$ is such that $\gamma^{ABCD}\gamma_D=b\gamma^{ABC}$. This solution is an analog of similar solutions appearing for membranes \cite{mt,Smolin, Moncrief, Hoppe}. It turns out however that equation (\ref{sq1}) gives (more) interesting results - a least when one concentrates on spherically symmetric case which will be now discussed.

\subsection{Spherically symmetric motion}

Let us now concentrate on the Friedmann-Robertson-Walker line element of spatially closed geometry
\begin{equation} \label{coord1}
g_{AB} dx^Adx^B = dt^2-a(t)^2 d\Omega^2, 
\end{equation}
\[
 d\Omega^2 = d\chi^2 + \sin^2 \chi( d\theta^2 + \sin^2 \theta d\varphi^2),
 \]
 \[
\chi \in [0, \pi], \theta \in [0, \pi], \varphi \in [0, 2\pi). 
\]
The corresponding expressions for $X^A$ for this geometry are \cite{rosen} 
\begin{equation} \label{coord2}
X^0=\int \sqrt{\dot{a}^2 + 1}dt, \ \ \  X^1 = a(t) \cos \chi, \ \ \ X^2 = a(t) \sin \chi \cos \theta, 
\end{equation}
\[
X^3 = a(t) \sin \chi \sin \theta \cos \varphi, \ \ \ X^4 = a(t) \sin \chi \sin \theta \sin \varphi 
\]
so that D = 5 - it is a minimal embedding. In order to find $a(t)$ let us substitute $X^A$ to the equations of motion (\ref{eom1}). We obtain
\begin{equation}  \label{eqa1}
3 + 3\dot{a}^2 + \ddot{a}a = 0 \ \ \Longrightarrow \ \ -a t''(a) + 3t'(a) + 3t'(a)^3 = 0   
\end{equation}
where $t=t(a)$ is the inverse function of $a=a(t)$. Integrating the last equation we obtain 
\[
t(a) = t_0 + \frac{A}{4} \ _2F_1\left(\frac{1}{2},\frac{2}{3}, \frac{5}{3}, A^2 a^4\right), \ \ \  A^2 = \frac{1}{a(t_0)^6(1+a(t_0)^2)} 
\]
where $_2F_1$ is the Gauss's hypergeometric function. An example of the solution 
for the initial conditions $a(0)=1$, $\dot{a}(0)=1$ is given in Figure 1. 
\begin{figure}[h]
\centering
\includegraphics[width=0.3\textwidth]{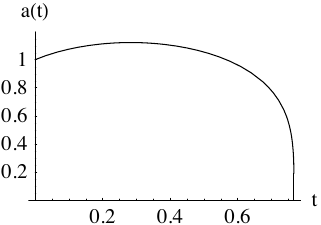}
\caption{Solution of equation (\ref{eqa1}) with  $a(0)=1$ and $\dot{a}(0)=1$.}
\end{figure}
We observe a typical behavior, just like in the case of membranes, that 
a 3-sphere will collapse to a point after a finite time.

\subsubsection{Dirac equation}

Let us now proceed to the quantum theory by linearizig 
the constraints. Because of the spherical symmetry there is only one dynamical variable $a(t)$. Therefore the quantum theory will be of quantum-mechanical type  i.e. we are looking for the equation involving derivatives w.r.t. $a$ - not the functional derivatives as in (\ref{sq1}) or (\ref{sq2}). This is completely analogous to considering mini-superspace models introduced by DeWitt \cite{DeWitt} where instead of a functional Wheeler-DeWitt equation one considers a  partial differential equation involving the scale factor. Because of that our starting point are not equations (\ref{sq1}), (\ref{sq2}) but the constraints (\ref{cons1}).

In the spherically symmetric case  the determinant $\det g_{rs}$ is simply $-a^6 \sin^4 \chi \sin^2 \theta$ hence the square root of (\ref{cons1}) can be written as 
\[
\gamma^A \mathcal{P}_A = -\lambda B a^3 \sin^2 \chi \sin \theta
\] 
where $B$ is a matrix s. t. $B^2 = \bold{1}$ (we choose $B = \bold{1}$), the minus sign 
is a convention. Introducing the canonical momenta $p_A$ for the whole 3-sphere
\[ 
p_A := \int_{t=const.} \mathcal{P}_A d\chi d \theta d \varphi 
\]
and going to the quantum theory via $p_A = -i\partial_A$, we find that 
\begin{equation} \label{dirace1}
( -i\gamma^A \partial_A|_a  + 2\pi^2 \lambda a^3)\psi = 0 
\end{equation}
where $|_a$ means taking the $a$-dependent part of the operator\footnote{Note that $\psi$ in (\ref{dirace1}) is a function of $t$ and $a$ and may also depend on the angles $\chi,\theta,\phi$ (c.p. (\ref{psi})). However we consider only such solutions for which $|\psi|$ is independent of $\chi$, $\theta$ and $\phi$ since we have already averaged the momenta.}. Therefore we have obtained a Dirac equation in $(4+1)$-dimensions with the cubic  scalar potential. 

Four comments are now in order. First, equation (\ref{dirace1}) is a mini-superspace counterpart of (\ref{sq1}), not (\ref{sq2}). The mini-superspace equation corresponding to (\ref{sq2}) would involve the integral of $M$ which turns out to be $0$ for  coordinates (\ref{coord2}). Therefore one obtains a continuous spectrum covering the whole real line in this case.

Second, note that the choice of the embedding is not unique since one could take higher dimensional Minkowski space. However one can always choose coordinates for which the momenta $\mathcal{P}_A=0$ for $A > 4$. For such coordinates the index $A$ in (\ref{cons1}) runs from $0$ to $4$ regardless of the embedding hence one arrives at (\ref{dirace1}) for any Minkowski embedding. Of course other types of geometries in general cannot be embedded in $(4+1)$ Minkowski space so that $D=5$ is not special for the generic case, however every Lorentzian 4-dimensional manifold can be locally embedded in $10$ dimensional Minkowski space \cite{theorem} - a result which can be useful for generalizations of the analysis presented below. 

Third, there is an ambiguity in interpreting $|\psi|^2$.  As a probability density of
\begin{itemize}
\item  a 3-sphere with the radius $\in [a, a + da]$ at time $t$
\begin{equation} \label{prob1}
p(a,t)= |\psi|^2
\end{equation}
\item  an infinitesimal element  $d^3\vec{x}$ of a 3-sphere being at distance $\in [a, a + da]$  from the origin of the coordinate system at time $t$- therefore the probability density for the whole 3-sphere would be
\begin{equation} \label{prob2}
p(a,t)= \frac{4}{3}\pi a^3|\psi|^2.
\end{equation}
\end{itemize}
It is in our opinion disputable which one should be chosen although the later seems more sensible since equation (\ref{dirace1}) is written in the spherical coordinates hence one expects the measure factor to appear (as it is the case in the Dirac or Schr\"odinger equation). We will discuss both (\ref{prob1}) and (\ref{prob2}) in this paper. 

Forth, from (\ref{dirace1}) one can consider the evolution of wave packets $\psi$, by using the Hamiltonian formulation 
\[
i\partial_t \psi = H \psi, \ \ \  H= \gamma^0(\gamma^k\partial_k|_a+2\pi^2 \lambda a^3), \ \ \ k=1,2,3,4.
\]
A detailed analysis of that dynamics will not be discussed in this paper.

The radial part of the Dirac operator in $4+1$ dimensions can be 
written in general as \cite{guma}
\begin{equation} \label{gumaeq}
\left( \partial_a  + \frac{K}{a}\right)G = (E + m -V)F, \ \ \  \left(-\partial_a + \frac{K}{a} \right)F = (E - m -V) G  
\end{equation}
where $E$ is the energy, $m$ is the mass term, $V$ is the potential (the $A_0$ component of the gauge field), $K=\pm(l+3/2)$, $l$ is the angular momentum (we shall consider $l=0$ from now on). The functions $F$ and $G$ are related to the wave function $\psi$ by
\begin{equation}  \label{psi}
\psi= a^{-3/2}e^{-iEt}\left[F(a) \phi_1(\chi,\theta,\varphi) + G(a) \phi_2(\chi,\theta,\varphi)\right]
\end{equation}
where $\phi_1$ and $\phi_2$ are suitably chosen, orthogonal, $a$ independent, spinors (for more details see \cite{guma}). We will normalize them so that
\[
|\psi|^2 = (F(a)^2+G(a)^2)/\frac{4}{3}\pi a^3.
\]
It is now clear that the choice of the interpretation of $|\psi|^2$ is crucial for the analysis of the probability density at $a=0$.

In our case we have $m= 2\pi^2 \lambda a^3$, $V=0$, therefore introducing dimensionless variables $x = a(2\pi^2 |\lambda|)^{1/4}$, $\epsilon = E/(2\pi 2|\lambda|)^{1/4}$ and the tension signature $\sigma = \lambda/|\lambda| = \pm 1$,  the spectral problem (\ref{gumaeq}) can be written in the matrix form
\begin{equation} \label{spectral}
h_{min} \phi = \epsilon_{min} \phi,  \ \ \ 
h_{min} :=  \left( \begin{array}{cc}
                     - \sigma x^3 & \partial_x + \frac{K}{x}    \\
                          -\partial_x + \frac{K}{x} &\sigma x^3
                              \end{array} \right).
\end{equation}
If $\phi^T=(F,G)$ is the eigen vector of $h_{min}$ for  $K=3/2$, $\lambda>0$ and some eigenvalue $\epsilon_{min}$ then $\phi^T = (G, F)$ 
solves $h_{min} \phi =-\epsilon_{min} \phi$ for $K = -3/2$, $\lambda>0$ and  $\phi^T = (-F,G)$ solves $h_{min} \phi =- \epsilon_{min} \phi$ for $K =3/2$, $\lambda<0$. Therefore we choose $K = 3/2$ and $\lambda>0$ from now on.

\subsubsection{Eigenfunctions at the origin}
Let us start with the following ansatz for functions $F$ and $G$
\begin{equation} \label{expansion}
F(x)=x^{s_1}e^{-x/2}(f_0 +f_1x +f_2x^2 + \ldots), \ \ \ f_0 \ne 0
\end{equation}
\[
G(x)=x^{s_2}e^{-x/2}(g_0 +g_1x +g_2x^2 + \ldots), \ \ \ g_0 \ne 0
\]
i.e. just like in the Dirac equation with the Coulomb potential (in which case one obtains $s_1=s_2$).
 Depending on the choice of the definition of the probability density one obtains different constraints on $F$ and $G$  -  either $F^2+G^2$ (cp. (\ref{prob2})) or $(F^2+G^2)/x^3$ (cp. (\ref{prob1})) must be normalizable. Clearly the second choice is more restrictive ($F$ and $G$ have to be at least such that  $(F^2+G^2)/x^3$ is finite at $x=0$) therefore to keep the discussion as general as possible we shall assume the first constraint. A class of functions obeying the first constraint should a priori be larger then for the second one. We shall now prove that a careful analysis of the spectral problem (\ref{spectral}) actually implies that these two classes are in  fact equal.
 
 First, let us use the fact that $F$ and $G$ can be expanded in the following orthonormal basis
 \begin{equation} \label{basis1}
e^{(\alpha)}_n(x) = \frac{1}{\sqrt{N_{n,\alpha}}} x^{\alpha} L^{(2\alpha)}_{n-1}(x)e^{-x/2}, 
\end{equation}
\[
N_{n,\alpha} = \frac{\Gamma(n+\alpha)}{(n+1)!},   \ \ \  \alpha=\min(s_1,s_2), \ \ \  n \in \mathbb{N}_+.
\]
where $L^{(\alpha)}_n(x)$ are generalized Laguerre polynomials. Clearly the condition $\int_0^{\infty} (F^2+G^2)dx < \infty$ implies that $\alpha >0$ however due to the $K/x$ term in the operator $h_{min}$  one in fact needs to assume that $\alpha \ge 1/2$ - otherwise the matrix representation of $h_{min}$ would not exist (i.e. the scalar products $(e_n,\frac{1}{x}e_m)=\int_0^{\infty}\frac{1}{x}e_ne_m dx$ would be infinite). This constraint can be further improved by using the expansion (\ref{expansion}). Because of the specific form of the operator $h_{min}$ (all powers of $x$ are integers) we may assume that $s_1-s_2 \in \mathbb{Z}$.  Substituting (\ref{expansion}) to (\ref{spectral}) and analyzing all the possibilities $s_1=s_2$, $s_1=s_2\pm 1$, $s_1=s_2 \pm 2$ et.c. one finds that there is only one possible choice where $\alpha>0$ and it is for  $s_2=s_1+1$ with  $s_1=K(=3/2)$. Therefore one should choose $\alpha=3/2$ which implies that not only  $F^2+G^2$ but also $(F^2+G^2)/x^3$ is normalizable.

Let us note again that the above conclusion is due to the term $K/x$ - both the singular behaviour and the coefficient $K$ are important. The term comes from the coordinate transformation of the kinetic part of the Dirac operator in (\ref{dirace1}) and is unremovable i.e. its generic form is $K/x$ with $K=\pm(2l+3)/2$ hence it persists for all angular momenta $l$. Analogous situation would not exist for the bosonic field. There the corresponding kinetic term would be the Laplace operator which in spherical coordinates has the centrifugal barrier term $l(l-d+2)/x^2$, in $d$ spatial dimensions. Therefore in the bosonic case there exist two sectors of the Hilbert space (for $l=0$ and for $l=d-2$) where there is no singularity.

We have just shown that  $|\psi|^2=(F(a)^2+G(a)^2)/a^3$ is regular therefore if one uses the definition (\ref{prob2}) then it follows that a 3-sphere cannot have a zero radius since the likelihood of that event is zero ($p(a=0,t_0)=0$). On the other hand our analysis also shows that $|\psi|^2$ is finite for $a=0$. Therefore a similar conclusion could not be obtained when taking (\ref{prob1}) as the definition of the probability density. 

Let us finally observe a close analogy between those results and the case of Dirac equation in the Coulomb field where  $|\psi|^2$ is in fact singular while  $F(a)^2+G(a)^2$ is zero for $a=0$.

\subsubsection{The spectrum}

The Hilbert space of the problem (\ref{spectral}) consists of square-integrable vectors $\phi^T=(F,G)$ on $[0,\infty)$ satisfying a condition $\phi(0) = 0$. Therefore the operator $\partial_x$ is antihermitian in this Hilbert space hence $h_{min}$ is hermitian which proves that the spectrum of $h_{min}$ is real.

Let us also observe that the spectrum of $h_{min}^2$ is discrete. To see this we use an identity 
\begin{equation}
h_{min}^2 = Q^2 +
\frac{3}{4x^2}\left( \begin{array}{cc}
                      1 & 0    \\
                      0 & 5
                              \end{array} \right), \ \ \ \
                      Q=         \left( \begin{array}{cc}
                -x^3 & \partial_x    \\
                     -\partial_x & x^3
                              \end{array} \right)
\end{equation}
hence we have an inequality $h_{min}^2 \ge  Q^2$. However $Q^2$ is discrete since introducing $a_{\pm} := F \pm G$ the eigen equation $Q^2\phi = \eta^2 \phi$ gives 
\[
h_{\pm}a_{\pm} = \eta^2 a_{\pm},  \ \ \ h_{\pm} := -\partial^2_x + x^6 \pm 3x^2,
\] 
i.e. $a_+$ and $a_-$ decouple and since $h_+$ and $h_-$ are discrete, $Q^2$ must also be discrete. Because $h_{min}^2$ is bounded from below by a discrete operator it follows that $h_{min}^2$ itself must be discrete \cite{reed} and therefore $h_{min}$ is discrete as well.

To find the exact spectrum of $h_{min}$ we use numerical methods. First, we calculate the representation of each operator in $h_{min}$, in some orthonormal basis. Our choice is (\ref{basis1}) for $\alpha=3/2$. Second we truncate that infinite matrix representation (i.e. truncate each matrix representation of operators appearing in the entries of the $2 \times 2$ matrix $h_{min}$) and then numerically diagonalize the resulting finite matrix. The spectra and the eigen vectors of the truncated matrices converge to their exact counterparts when the size of 
the matrix is increased. The results of this numerical approach are presented in Figure 2. 
\begin{figure}[h]
\centering
\includegraphics[width=0.7\textwidth]{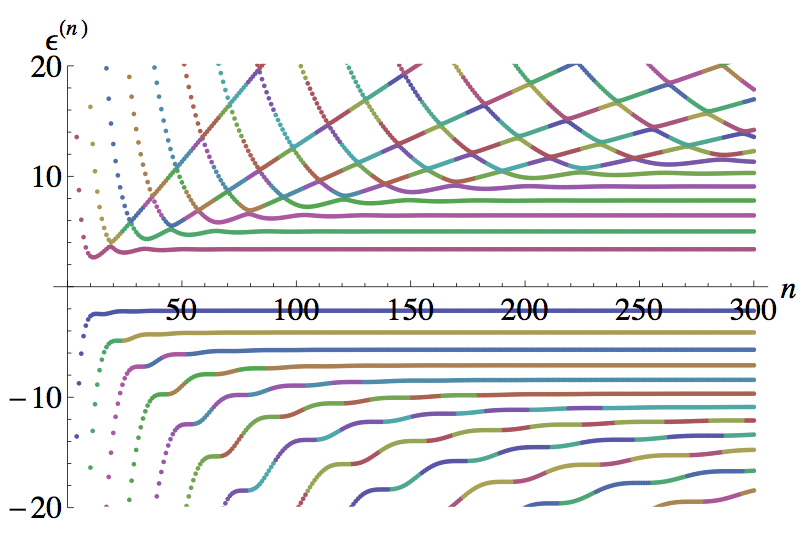}
\caption{Convergence of energy levels of $h_{min}$ (for $K=3/2$, $\sigma=1$) with  $n$ - the size of truncated matrix representation - up to $n = 300$.}
\end{figure}
It follows that the first positive and negative energy levels are 
\[
\epsilon \approx  3.4, \  5.0, \  6.5, \  7.8, \  9.0, \  10.3, \  11.4, \ldots, 
\]
\[
-\epsilon \approx 2.2, \  4.1, \ 5.7, \ 7.1, \ 8.4, \ 9.7, \ 10.9, \ldots \ . 
\]
The plots for $F^2(x)+G^2(x)$ are presented in Figure 3.
\begin{figure}[h]
\centering
\includegraphics[width=1\textwidth]{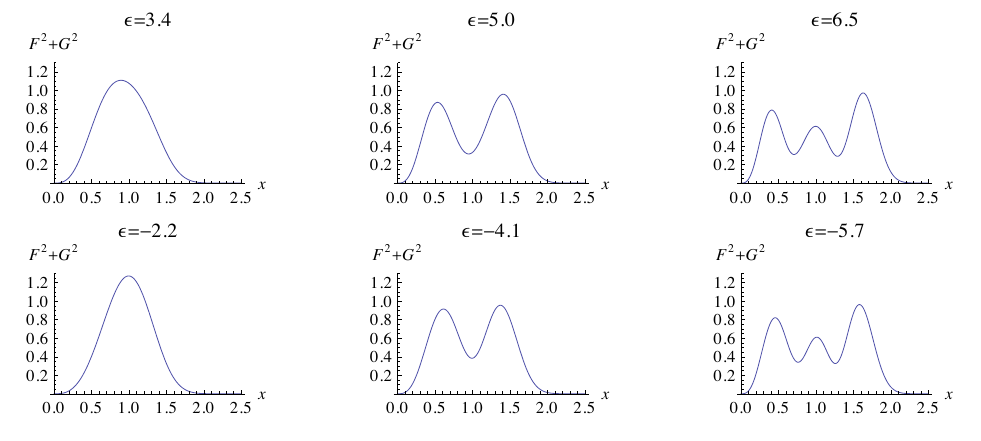}
\caption{$F^2(x)+G^2(x)$ for first three positive and negative energy states of $h_{min}$.}
\end{figure}
The negative energy solutions are normalizable therefore one cannot ignore them. This would be troublesome if the 3-sphere was interacting with some external field or with another membrane in the embedding space $E$ - in which case a 3-sphere could loose energy. Because the spectrum is not bounded from below this would imply that the energy of a 3-sphere could become arbitrary negative. Consistency would require that there is no such external fields and that there is only one 3-sphere in $E$. Otherwise one needs to introduce a Dirac sea (of negative energy 3-spheres) i.e. assume that all the negative energy states are filled.

\section{General action}
In this section we will generalize  previous considerations to the case of Einstein-Hilbert action with the cosmological constant term and matter. Therefore we consider
\begin{equation} \label{action2}
S= S_{\lambda} +S_{EH}+S_{m}
\end{equation}
\[
S_{\lambda}= -\lambda \int_{\mathcal{M}} \sqrt{-g} \ d^4x \ \ \ \ S_{EH} =-\frac{1}{2\kappa}\int_{\mathcal{M}} \sqrt{-g}R \ d^4x,
\]
where  $\kappa=8\pi G$, $R$ is the Ricci scalar, $S_m$ is the action for matter \footnote{We shall not consider, in this paper, generalizations of the Einstein-Hilbert action to higher order Largangians such as $f(R)$ gravity. Quite surprisingly these models exhibit a bifurcation into almost Einsteinian branches \cite{schnuck} which provides a different point of view on $S_{EH}$ and hence the singularity problem. }.  The tension $\lambda$ is related to the cosmological constant $\Lambda$ via $\lambda = \Lambda /8\pi G$, from now on we will set $G=1$. Our view on the action (\ref{action2}) is non-standard here. Usually one takes the Einstein-Hilbert term as a starting point and only then adds "something else" e.g. $S_{\lambda}$ or $S_m$. Here we consider space-time as embedded manifold with tension $\lambda$ therefore $S_{\lambda}$ is a starting point while $S_{EH}$ and $S_{m}$ are in addition to it. Moreover the Einstein-Hilbert term is defined on $\mathcal{M}$ not in the embedding space $E$. As such $S$ can be also viewed as an action for the vector field $X^A(x)$ on $\mathcal{M}$ - with no reference to the embedding.

 Varying $S$ with respect to the $X^A$ one obtains
\[
\delta_X S = \frac{1}{2}\int_{\mathcal{M}} \left(-\lambda g^{\mu\nu} + \frac{1}{\kappa} G^{\mu\nu}- T^{\mu\nu}\right)\delta_X g_{\mu\nu} \sqrt{-g}d^4x
\]
where $G_{\mu\nu}= R_{\mu\nu}-\frac{1}{2}g_{\mu\nu}R$ is the Einstein tensor, $R_{\mu\nu}$ in the Ricci tensor ($R_{\mu\nu}:= R^{\alpha}_{\mu\alpha\nu}$) and $T_{\mu\nu}:=2\delta S_m/ \delta g^{\mu\nu}$ is the energy momentum tensor. The variation of $g_{\mu\nu}$ gives $2 \partial_{\mu} X_A \delta \partial_{\nu}X^A$ hence the equations of motion are
\begin{equation}  \label{eom2}
\partial_{\mu}\left[  \sqrt{-g}\left(-\lambda g^{\mu\nu} + \frac{1}{\kappa} G^{\mu\nu} - T^{\mu\nu}\right)\partial_{\nu} X^A\right] =0.
\end{equation}
These equations can be obtained form the usual Euler-Lagrange equations \cite{pavsic3,pavsic4}. 

\subsection{Spherically symmetric motion}

If $g_{\mu\nu}$ satisfies Einstein's equations with the cosmological constant and matter terms
\begin{equation} \label{einstein}
-\lambda g^{\mu\nu} + \frac{1}{\kappa} G^{\mu\nu} =  T^{\mu\nu}
\end{equation}
then clearly equation (\ref{eom2}) is satisfied however it is not so clear if there exist coordinates $X^A$ which give rise to that metric via (\ref{embed}). The problem lies within the energy-momentum tensor term - if $T_{\mu\nu}=0$ then one can always find (locally) the embedding coordinates however if $T_{\mu\nu} \ne 0$ such coordinates may not exist. We shall now verify this remark for the Friedmann-Robertson-Walker line element (\ref{coord1}) and the energy-momentum tensor for a perfect fluid
\begin{equation} \label{emtensor}
T_{\mu\nu}=(\rho + p)\delta_{\mu 0} \delta_{\nu 0}-pg_{\mu\nu}.
\end{equation}
There are a priori five equations in (\ref{eom2}) for each $A$ however for $A=1,2,3,4$ one obtains the same equation which is
\begin{equation}  \label{eomf}
N:=3 (1 + \dot{a}^2)^2 - 3\kappa a^2 (\lambda - p + (\lambda + \rho) \dot{a}^2) -\kappa (\lambda + \rho) a^3 \ddot{a} +  9a(1+\dot{a}^2) \ddot{a}=0 
\end{equation}
while for $A=0$ we find
\[
N -3\kappa a^2 (\rho + p) =0.
\]
It follows that nontrivial solutions for $a$ exist only if we put $\rho+p=0$ hence it is impossible to find embedding coordinates for for FRW metric when $\rho>0$ and $p>0$. 
On the other hand if we allow the pressure to be negative then the condition $\rho +p=0$ can be obtained  by considering the matter term $S_m$ for the scalar field $\phi$ in a particular potential $V(\phi)$. Then one can show that the equation of state $\rho + p =0$ is approximately satisfied for slowly rolling fields (i.e. $\frac{1}{2}\dot{\phi}<< V(\phi)$) \cite{inf1,inf2,inf3}. 
Let us therefore assume that $p=-\rho$. Equation (\ref{eomf}) can be written in a convenient form
\begin{equation}  \label{eqa2}
3+3\dot{a}^2+a\ddot{a}=\frac{24(1+\dot{a}^2)^2}{9-\kappa (\lambda +\rho)a^2 + 9\dot{a}^2}
\end{equation} 
 and we see that in the limit $\kappa \to \infty$ one recovers the equation for minimal 4-volumes (\ref{eqa1}). It is now straightforward to verify that (\ref{eqa2}) is a consequence of Firedmann equations for $\rho+p=0$. Substituting (\ref{emtensor}) and (\ref{coord1}) to (\ref{einstein}) we obtain standard equations
\begin{equation} \label{fried}
\left(\frac{\dot{a}}{a}\right)^2 = -\frac{1}{a^2}+\frac{\kappa (\lambda+\rho)}{3}, \ \  \ \ \  \frac{\ddot{a}}{a} = \frac{1}{3}\kappa (\lambda+\rho).
\end{equation}
Using expressions for $\dot{a}$ and $\ddot{a}$ obtained from (\ref{fried}) we find that the l.h.s. and the r.h.s. of (\ref{eqa2}) coincide (giving $4\kappa(\lambda+\rho) a^2/3$ each).
Therefore we have shown that (\ref{fried}) imply (\ref{eqa2}).  It is reasonable to ask if the converse is also true i.e. whether (\ref{eqa2}) are equivalent to (\ref{fried}). 
We will now show that this is not true i.e. the space of solutions of (\ref{eqa2}) is much larger then the space of solutions of (\ref{fried}). 

For simplicity let us assume that $\rho=p=0$ i.e. the Universe is empty. Then the general solution of the Friedmann equations (\ref{fried}) for $\lambda>0$ are non singular 
\begin{equation} \label{dssol}
a(t)=\sqrt{3/\lambda\kappa}\cosh(t\sqrt{\lambda \kappa/3})
\end{equation}
(De Sitter solution of spatially closed Universe) while for negative $\lambda$ and $\rho=0$ equations (\ref{fried}) have no solutions. As shown above (\ref{dssol}) is also the solution of (\ref{eqa2}) however it is not clear if it is a general solution of (\ref{eqa2}) \footnote{This is not in conflict with the uniqueness of solutions since the space of boundary conditions for (\ref{eqa2}) is much larger then the space of boundary conditions of (\ref{fried}). A simple example of this sort is a pair of equations $\dot{a}=1$, $a=t$ which have general solution $a(t)=t$. On the other hand they also imply that $\dot{a}a=t$ the general solution of which is $a(t)=\sqrt{t^2+A}$.}. 

Let us first analyze the numerical solutions of (\ref{eqa2}) for $\rho=0$ - Figure 4.
\begin{figure}[h]
\centering
\includegraphics[width=0.5\textwidth]{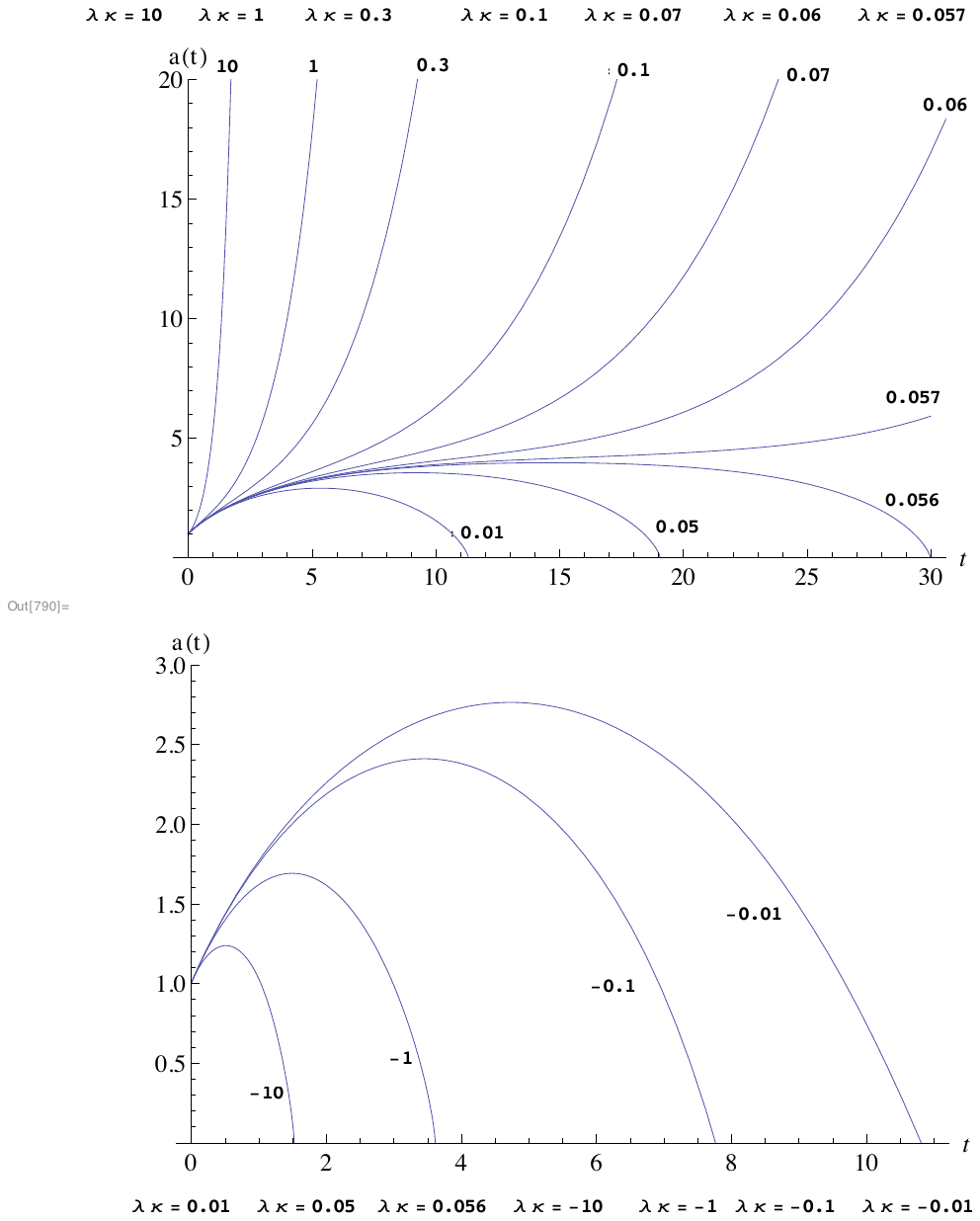}
\caption{A family of solutions of (\ref{eqa2}) with $a(0)=1$, $\dot{a}(0)=1$ for various (positive and negative) values of $\lambda \kappa$. There exists a critical value below which the solutions develop singularity.}
\end{figure}
We see that not only the solutions do not resemble (\ref{dssol}) but also that there is a critical value of $\lambda \kappa$ below which the solutions develop singularity in finite time (for the boundary conditions $a(0)=1$, $\dot{a}(0)=1$ the critical value satisfies the bounds  $0.056<\lambda_0 \kappa <0.057 $). This result is rather interesting since we obtained  three classes of solutions (asymptotically expanding, asymptotically static or contracting) similar to Friedmann solutions however let us point out that unlike for Friedman solutions there is no matter here. Therefore the embedding equations (\ref{eom2}) allow for solution which do not solve the Einstein equations - in particular they allow for singular solutions for $T_{\mu\nu}=0$.

Let us now give more rigorous proof that solutions of (\ref{eqa2}) are more general then (\ref{dssol}). It is useful to introduce the function $F$ defined as ($\rho=0$)
\begin{equation} \label{step1}
F:= \frac{3}{\kappa \lambda}\frac{\dot{a}^2+1}{a^2}.
\end{equation}
 In case of Friedmann equations (\ref{fried}) $F$ is simply 1 - we will now be looking for more general $F$ of the form $F=F(a)$. 
Expressing $\dot{a}$ in terms of $F$ and $a$ and substituting the result to (\ref{eqa2}) we obtain
\begin{equation} \label{step2}
\ddot{a} = \frac{\kappa\lambda}{3}\frac{(1-F/3)aF}{F-1/3}.
\end{equation}  
Equations (\ref{step1}) and (\ref{step2}) are equivalent to (\ref{eqa2}) for $\rho=0$.
Using the formula $\ddot{a}=\frac{1}{2}\frac{d \dot{a}^2}{da}$ and applying it to (\ref{step2}) we obtain
\begin{equation} \label{step3}
\dot{a}^2 = \frac{2}{3}\kappa \lambda \int_{0}^{a} \frac{\tilde{a} F(\tilde{a})(1-F(\tilde{a})/3)}{F(\tilde{a})-1/3}d \tilde{a}.
\end{equation}
On the other hand $\dot{a}^2$ from (\ref{step1}) is
\begin{equation} \label{step4}
\dot{a}^2 = \frac{\kappa \lambda}{3}a^2 F -1.
\end{equation}
Equating (\ref{step3}) with (\ref{step4}) we obtain an integral equation for $F$. The derivative w.r.t. $a$ of that equation gives
\[
a F'(a) = 8F(a)\frac{1-F(a)}{3F(a)-1}.
\]
Integrating the above equation we obtain a non-linear equation for $F$
\begin{equation}  \label{step5}
(F-1)^2F = e^A/a^8, \ \ \ \ A \in \mathbb{R}.
\end{equation}
We now see that the Friedmann equations case, $F=1$, corresponds to taking the limit $A \to -\infty$ in (\ref{step5}). However solutions of (\ref{step5}) are more general then that. Equation (\ref{step5}) has three roots, two of which are complex and one is real which is
\[
F(a) = \frac{1}{3}(2+x+1/x), \ \ \ x= \left( \frac{27e^A+3\sqrt{3}\sqrt{27e^{2A}-4a^8 e^A}}{a^8} -1 \right)^{1/3}
\]
To find the solution for $a(t)$ one can use the definition (\ref{step1}) to arrive at the complicated integral
\[
\int_{a_0}^a \frac{d\tilde{a}}{\sqrt{\frac{\kappa \lambda}{3}F(\tilde{a})\tilde{a}^2-1}} = t-t_0, \ \ \ a_0:=a(t_0).
\]
This completes the proof that a class of solutions of (\ref{eqa2}) is larger then a class of solutions of (\ref{fried}).

\subsection{Dirac equation}

In this section we shall discuss the singularity formation in the context of the Dirac equation using similar method as in Section 2. For general coordinates the canonical momenta for the action (\ref{action2}) are now
\[ 
\Pi_A = \mathcal{P}_A+  \sqrt{-g}\left(\frac{1}{\kappa}G^{0\mu} - T^{0\mu}\right)\partial_{\mu} X_A
\]
where $\mathcal{P}_A$ are as in (\ref{momenta1}). The constraint (\ref{cons1}) for $\mathcal{P}_A$ still holds therefore its linearized form will be
\begin{equation} \label{dirace2}
\gamma^A \left[ \Pi_A -  \sqrt{-g}\left(\frac{1}{\kappa}G^{0\mu} - T^{0\mu}\right)\partial_{\mu} X_A\right]\Psi=\lambda \sqrt{-\det g_{rs}}.
\end{equation}
To proceed further we now concentrate on the Friedmann-Robertson-Walker line element for closed 3-geometry. Let us therefore assume that the metric is given by (\ref{coord1}) and that there is no matter at all (as shown previously this case contains singular solutions).  Following Section 2 we introduce the average momenta for the whole 3-sphere
\[
\pi_A= \int_{t=const.} \Pi_A d\chi d \theta d\varphi
\] 
and, as in (\ref{dirace1}), we will consider the wave equation by substituting $\pi_A\to -i\partial_A$. Equation (\ref{dirace2}) becomes now
\begin{equation} \label{dirace3}
( -i\gamma^A \partial_A|_a  + 2\pi^2 \lambda a^3)\psi  + \gamma^0 \hat{V}\psi = 0 
\end{equation}
where $\hat{V}$ is an operator corresponding to the classical expression
\[
V= -\frac{1}{\kappa}\int G^{00}\sqrt{-g}d\chi d \theta d\varphi = -\frac{6\pi^2}{\kappa}(\dot{a}^2a+a)
\]
where we used the fact that $G^{\mu\nu}$ is diagonal and that the integral over $\partial_0 X_A$ is non zero only if $A=0$ (moreover $G^{00}=3(\dot{a}^2+1)/a^2$ and $\sqrt{-g}=a^3\sin^2 \chi \sin \theta$). Note that $\hat{V}$ enters equation (\ref{dirace3}) in the same way the zero'th component of the electromagnetic field enters the Dirac equation with an external field. To find the quantum counterpart of the term involving the time  derivatives of $a$ we first calculate the momenta conjugate to $a$
\[
p_a =-\frac{3\pi}{2}\dot{a}a, \ \ \ V=-\frac{1}{3\pi}\frac{p_a^2}{a} - \frac{3\pi}{4}a.
\] 
The momenta are calculated using a standard method \cite{rev1} (in deriving $p_a$ one uses the action (\ref{action2}) augmented by a boundary term needed for the consistent variational procedure - that term does not affect the equations of motion nor the constraints). Second, there is an ambiguity in choosing the differential operator corresponding to $p_a^2/a$ term. We choose the ordering following DeWitt \cite{DeWitt} as
\[
\frac{p_a^2}{a}  \to -\frac{1}{a^{1/4}}\partial_a  \frac{1}{a^{1/2}} \partial_a  \frac{1}{a^{1/4}}.
\]
Introducing dimensionless units as in previous section we find that the eigenvalues of (\ref{dirace3}) can be obtained from the following spectral problem
\begin{equation} \label{des}
h_{dS}\phi= \epsilon_{dS} \phi, \ \ \  h_{dS}=h_{min} + v(x)\bold{1},
\end{equation}
\[
v(x) = \frac{1}{2\beta}\Delta - \frac{1}{2}\beta x, \ \ \ \ \Delta = \frac{1}{x^{1/4}}\partial_x  \frac{1}{x^{1/2}} \partial_x  \frac{1}{x^{1/4}}
\]
where $h_{min}$ is as in (\ref{spectral}) and $\beta=3\sqrt{\pi/\Lambda}$ is a dimensionless constant. If we were to substitute the experimental value of $\Lambda$ then the term proportional to $\Delta$ would be extremely small while the linear term  would be very large and negative.

\subsubsection*{Singularity avoidance}

As in Section 2 we shall now be looking for functions $F$ and $G$ of the form (\ref{expansion}). The operator $\Delta$ contains terms $x^{-3}$, $x^{-2}\partial_x$ and $x^{-1}\partial_x^2$ which imply that one should take $\alpha:=\min(s_1,s_2) \ge 3/2$ in order to have the scalar products $(e_n,\Delta e_m)=\int_0^{\infty}e_n 
\Delta e_m dx$ finite. Using the same analysis as before we find that this condition can be satisfied this time only when $s_1=s_2$ in which case one finds that the equation for (say) $s_1$ is $s_1^2 + s_1 +7/16=0$. The roots are  $1/4$ and $7/4$ but $\alpha \ge 3/2$ hence we take $\alpha=7/4$ which proves that not only the probability density (\ref{prob2}) but also (\ref{prob1}) is zero for $a=0$, concretely
\[
|\psi|^2 = (F(x)^2+G(x)^2)/\frac{4}{3}\pi x^3 = \sqrt{x}e^{-x/2}r(x)
\]
where $r(x)$ is regular at $x=0$.  Clearly, this result is much stronger then the one obtained in Section 2 -  it shows that the eigen functions are $0$ for $a=0$. 

The Hamiltonian formulation is now
\[
i\partial_t \psi = H \psi, \ \ \  H= \gamma^0(\gamma^k\partial_k|_a+2\pi^2 \lambda a^3)-\hat{V}, \ \ \ k=1,2,3,4
\]
where the wave packets $\psi$'s can be expanded in terms of eigen functions  of which we know that are 0 for $a=0$. Therefore, no matter which definition of the probability density one chooses ((\ref{prob1}) or (\ref{prob2})), one concludes that the 3-sphere cannot have zero radius i.e. the likelihood of that event is 0. This certainly fulfills the DeWitt's requirement and hence we conclude that the singularity is not present. 

A contracting 3-sphere will approach a minimal size and then start expanding. It is likely that this expansion will then be slowed down and stopped due to the $x^3$ terms in $h_{dS}$ therefore the evolution of the wave packet would be cyclic. To verify this remark one would have to perform the corresponding numerical analysis.

\subsubsection*{The spectrum}

The operator $v(x)$ is negative definite and it enters $h_{dS}$ with the plus sign therefore the eigenvalues of $h_{dS}$  will be lowered compared to the eigenvalues of $h_{min}$.
Moreover $h_{dS}$ contains a parameter $\beta$ which influences the spectrum considerably - that parameter cannot be eliminated by some rescaling of coordinate $x$.  Several spectral lines for $\beta=10$ and $\beta=20$ are presented in Figure 5.  \\

\begin{figure}[h]
\centering
\includegraphics[width=1\textwidth]{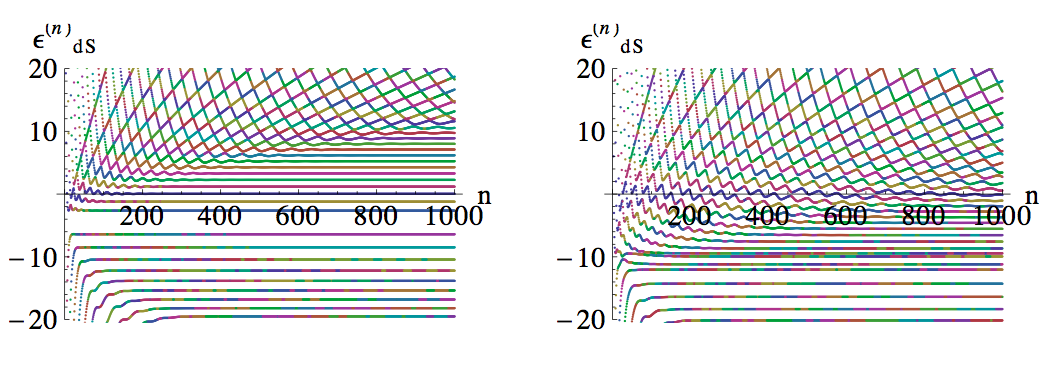}
\caption{Spectral lines of $h_{dS}$ for $\beta=10$ (left) and $\beta=20$ (right). The cutoff reached is $n=1000$. To obtain reliable positive spectral values for $\beta=20$ one would have to go to higher $n$. On the other hand negative eigenvalues converge much quicker.}
\end{figure} 

\noindent We observe a discrete spectrum. The positive and negative eigenvalues obtained for $\beta=10$ are
\[
\epsilon \approx 0.04, \ \  1.2, \ \ 2.2, \ \ 3.3, \ \ 4.3, \ \ 5.2, \ldots,
\]
\[
-\epsilon \approx 1.2, \ \  2.6, \ \  6.4, \ \ 8.5, \ \ 10.4, \ \  12.2, \ldots \  .
\]
The eigenvalues, compared to the spectrum of $h_{min}$ in (\ref{spectral}), are shifted by (roughly) $- \beta/2$. Therefore the lowest positive energy state 
will now have several "bumps" instead of just one bump (see Figure 6).
\begin{figure}[h]
\centering
\includegraphics[width=1\textwidth]{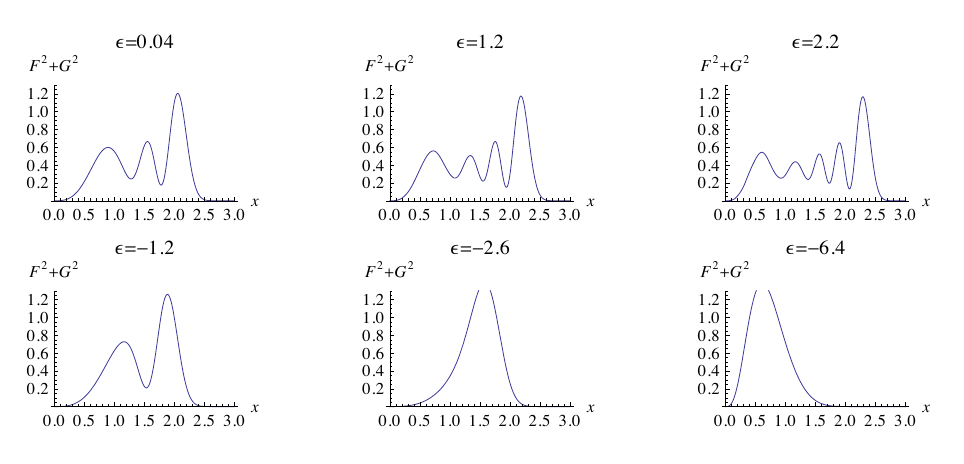}
\caption{$F^2(x)+G^2(x)$ for first three positive and negative energy states of $h_{dS}$ for $\beta=10$.}
\end{figure}

 The energies are calculated here in units of $(2\pi^2 \lambda)^{1/4}=(\pi \Lambda/4)^{1/4}$ therefore  the density of spectral lines of $h_{dS}$ and $h_{min}$ is very large if one substitutes  $\beta \sim 10^{60}$ (i.e. $\Lambda \sim 10^{-120}$). However the appearance of $\beta$ in $h_{dS}$ makes the spectrum even more denser (compared to $h_{min}$) so  that practically any energy is allowable as if the spectrum was continuous. Therefore the smallness of the cosmological constant implies in this model that the empty Universe may have arbitrary energy.  On the other hand if $\Lambda$ was say $\approx 1$ the quantisation of the energies of the Universe would be noticeable.

\section{Summary}
By considering spacetime $\mathcal{M}$ as as an embedded Lorenzian 4-manifold is some ambient, higher-dimensional space $E$ one necessarily faces a question whether $E$ is in any sense physical. In this paper we make no arbitrary statements concerning this problem however we do take advantage of the fact that certain, conceptual problems appearing in the formulation of Quantum Gravity by considering  Wheeler-DeWitt equation (e.g. the problem of time) are naturally solved within this approach.  Moreover the constraints (\ref{cons1}) appearing in such approach are very similar, in the form, to the mass-shell constraint for  point like particles and therefore one is tempted to apply the Dirac's "square-root" procedure also in this case. By doing so one arrives at the functional Dirac-like equation for embedded manifolds which are linear in momenta. There are at least two distinct ways of doing this  ((\ref{sq1}) and (\ref{sq2})) and in our opinion both are important.

 We then concentrate on the spherically symmetric motions by considering $S^3$ with one dynamical variable - $a(t)$ - and make a proposal (based on (\ref{sq2})) for a quantum mechanical system behind such mini-superspace model. For minimal 4-volumes we find that the spectrum of the corresponding Dirac equation (\ref{dirace1}) is real and discrete - it contains positive and  negative eigenvalues which are not bounded form below.  By careful analysis of the wave functions at the origin we showed that the amplitude $|\psi|^2$ is finite at $a=0$ hence it is disputable whether a 3-sphere can collapse due to the ambiguity of defining the probability density. 
 
 It turns out that considering general action with the Einstein-Hilbert term one can improve the situation by finding that the amplitude  $|\psi|^2=0$ for $a=0$. It is therefore in accordance with DeWitt's boundary condition that the function vanishes for those classical configurations where the singularity appears. This analysis is performed for the Friedmann-Robertson-Walker metric for closed spatial geometries. One would like to call this model an embedded closed Friedmann model with no matter (or embedded closed De Sitter model) however it is important to note that the classical equations of motion of the embedding picture (\ref{eqa2}) are much more general then Friedmann equations (\ref{fried}). In particular they allow for singular solutions for both positive and negative tension $\lambda=\Lambda/\kappa$ which is not the case in (\ref{fried}) for $\rho=0$.
We have shown that in the quantum picture, when one considers the Dirac equation (\ref{dirace3}), these singularities are avoided. 
 
 To generalize this approach one should consider the case $\rho >0$ which can be done by introducing slowly rolling scalar fields.

\section{Acknowledgements}
I thank J. Hoppe, J.-L. Lehners and  D. Lundholm for discussions. This work was supported by DFG (German Science Foundation) via the SFB grant.

\end{document}